\documentclass[11pt]{article}

\def\Re{\mathop{\rm Re}\nolimits}
\def\Im{\mathop{\rm Im}\nolimits}

\def\rmi{{\rm i}}
\def\rmd{{\rm d}}
\usepackage{times}
\usepackage{amsmath}
\usepackage{graphicx}
\usepackage{latexsym}
\usepackage{amsfonts}
\usepackage{amssymb}
\usepackage{epsfig}
\usepackage{pstricks}

\makeatletter
\@addtoreset{equation}{section}
\makeatother

\newcommand{\beqn}{\begin{eqnarray}}
\newcommand{\eeqn}{\end{eqnarray}}
\newcommand{\be}{\begin{equation}}
\newcommand{\ee}{\end{equation}}

\newcommand{\ft}[2]{{\textstyle {\frac{#1}{#2}} }}

\def\be{\begin{equation}}
\def\ee{\end{equation}}
\def\bea{\begin{eqnarray}}
\def\eea{\end{eqnarray}}

\newskip\humongous \humongous=0pt plus 1000pt minus 1000pt

\newif\ifdtup

\begin{document}


\begin{titlepage}
\begin{flushright}
arXiv:0801.1666\\
MPP-2008-7\\
\end{flushright}
\vspace{.5cm}
\begin{center}
\baselineskip=16pt {\bf \LARGE  Generalized Chern-Simons Terms and   }
\vskip4mm  {\bf \LARGE  Chiral Anomalies in $\mathcal{N}=1$ Supersymmetry }\\
\vspace*{7mm} \vfill
 {\Large
Marco Zagermann
} \\
 \vspace*{5mm} \vfill
 {\small
   Max-Planck-Institut f\"{u}r Physik, F\"{o}hringer Ring 6,\\
 80805 Munich,  Germany \\ \vskip4mm  }
\end{center}
 \vfill
 \begin{center}
 {\bf Abstract}
 \end{center}
 {
The gauging of axionic, Stueckelberg-type, shift symmetries with generalized Chern-Simons terms and the Green-Schwarz mechanism of anomaly cancellation has recently been studied
in the context of certain string compactifications and with regard to the phenomenology of unusual variants of $Z^{\prime}$-bosons. In this talk, following recent work with De Rydt, Rosseel, Schmidt and Van Proeyen, I summarize how such theories can
be reconciled with global and local $\mathcal{N}=1$ supersymmetry. Two simple consistency conditions are shown to encode strong constraints on the allowed
anomalies for different types of gauge groups. 
}\vspace{2mm} \vfill \hrule width 3.cm
{\footnotesize \noindent e-mail: zagerman@mppmu.mpg.de }
\end{titlepage}
\newpage
\section{Introduction}
In generic low energy effective field theories, gauge fields appear with non-minmal kinetic terms in which the field strengths may multiply scalar field dependent coefficients. In the context of $\mathcal{N}=1$ supersymmetry, these scalar field dependent coefficients can be  expressed in terms of the holomorphic gauge kinetic function, $f_{AB}(z^{i})$, that depends on the complex scalars, $z^{i}$, of the chiral multiplets \footnote{We use
 the non-Abelian field strength $
  {\cal F}_{\mu \nu }^A = F_{\mu \nu }^A + W_\mu ^B W_\nu ^Cf_{BC}{}^A$, where $F_{\mu \nu }^A=
 2\partial _{[\mu }W_{\nu ]}^A$ is the Abelian part and $f_{AB}{}^{C}$ denotes the structure constants of the gauge group, $G$, with $A,B,C, \ldots = 1,\ldots,\dim G$. The tilde denotes the
 Hodge dual, $\tilde{F}^{\mu\nu}=-\ft12ie^{-1}\epsilon^{\mu\nu\rho\sigma}F_{\rho\sigma}$, and $e$ is the vierbein determinant.}:
\begin{equation}
  e^{-1}{\cal L}_1  =
   -\ft14\Re f_{AB} {\cal F}_{\mu \nu }^A {\cal F}^{\mu \nu \,B} +\ft14\rmi\Im f_{AB} {\cal F}_{\mu \nu }^A \tilde {\cal F}^{\mu \nu
   \,B} \, .
 \label{L1f}
\end{equation}
If some of the scalars $z^{i}$ are charged under the gauge group $G$, they may introduce a non-trivial gauge transformation of the gauge kinetic function,
 $\delta(\Lambda)  f_{AB}= \Lambda ^C\delta _C f_{AB}\neq 0$, where $\Lambda^{C}(x)$ 
is the gauge parameter. Gauge invariance of (\ref{L1f}) then requires that this induced transformation be of the form
 \begin{equation}
 \delta _C
f_{AB}=f_{CA}{}^D f_{BD} + f_{CB}{}^Df_{AD}\, . \label{conditionfinvC1}
\end{equation}
This, however, is only the most general form when we consider the gauge invariance of the kinetic term (\ref{L1f}) in isolation. If we take into account also other possible terms in the classical and quantum effective action, we can allow
for a slightly more general transformation property of $f_{AB}$:
\begin{equation}
\delta _C f_{AB}=\rmi C_{AB,C} +f_{CA}{}^D f_{BD} + f_{CB}{}^Df_{AD}\,.
\label{conditionfinvCb}
\end{equation}
Here, $C_{AB,C}$ is a constant real tensor symmetric in the first two
indices. One can show  (see e.g.  \cite{DeRydt:2007vg})  that the above transformation law is of the most general 
form consistent with symplectic duality invariance of the action and that the closure of the
gauge algebra imposes the constraint
\begin{equation}
C_{AB,E}f_{CD}{}^E -2  C_{AE,[C}f_{D]B}{}^E -2  C_{BE,[C}f_{D]A}{}^E= 0
\, .
 \label{NonAbelianCident}
\end{equation}
Transformations with non-vanishing $C_{AB,C}$ are called axionic shift
symmetries. The simplest example is given by $f_{AB}(z)=z\delta_{AB}$
with a scalar field $z$ that transforms as $\delta_{C}z=ia_{C}$, for some constants $a_{C}$, leading to $C_{AB,C}=\delta_{AB}a_{C}$. 

Obviously, (\ref{L1f}) is not invariant under (\ref{conditionfinvCb}):
\begin{equation}
\delta(\Lambda ) e^{-1}{\cal L}_1  = \ft 14 \rmi C_{A B,C} \Lambda ^C
 {\cal F}_{\mu \nu }^A \tilde {\cal F}^{\mu \nu \,B }\,.
 \label{delLambdaSf1}
\end{equation}
There are two known mechanisms that can cancel non-invariances of the form
(\ref{delLambdaSf1}): (i) Generalized Chern-Simons terms (GCS terms) and (ii) Quantum anomalies. To understand this, we split the tensor
\begin{equation}
C_{AB,C}=C_{AB,C}^{(s)}+C_{AB,C}^{(m)}
\end{equation}
into its completely symmetric part, $C_{AB,C}^{(s)}=C_{(AB,C)}$, and a part of mixed symmetry, $C_{AB,C}^{(m)}$, with $C_{(AB,C)}^{(m)}=0$. \\

\noindent\textbf{(i) Generalized Chern-Simons terms}\\
As was first found in the context of $\mathcal{N}=2$ supergravity in \cite{Sugra}, and discussed more generally in \cite{deWit:1987ph}, 
generalized Chern-Simons terms of the form
\begin{equation}
  S_{\rm CS}=\int d^4 x  C^{\rm (CS)}_{AB,C}\varepsilon ^{\mu \nu \rho \sigma }
  \left(\ft16 W_\mu ^CW_\nu ^A F_{\rho \sigma }^B + \ft18 f_{DE}{}^A W_\mu ^DW_\nu ^E W_\rho ^CW_\sigma
  ^B\right)
 \label{SCS1}
\end{equation} 
can cancel the terms proportional to the mixed part, $C_{AB,C}^{(m)}$, in the variation (\ref{delLambdaSf1}), provided we identify $C_{AB,C}^{\rm (CS)}=C_{AB,C}^{(m)}$.\\

\noindent\textbf{(ii) Quantum anomalies}\\
An anomalous spectrum of chiral fermions induces a gauge non-invariance of the quantum effective action, $\Gamma[W_{\mu}^{A}]$, of the form $
\delta(\Lambda)\Gamma[W_{\mu}^{A}]=\int d^4 x \Lambda^{C} \mathcal{A}_{C}, $
where $\mathcal{A}_{C}$ denotes the consistent anomaly
\begin{equation}
{\cal A}_C=-\frac{i}{4}  
 \left[  d_{ABC} F_{\mu\nu}^B
+\left( d_{ABD}f_{CE}{}^B +\ft32 d_{ABC}f_{DE}{}^B \right)
W_\mu^DW_\nu^E\right]
\tilde F^{\mu\nu A} \label{gaugeanom}
\end{equation}
with
$d_{ABC}\sim \textrm{Tr}(\{T_{A},T_{B}\}T_{C})$,
where $T_{A}$ are the  gauge generators \footnote{The form of the anomaly depends on the renormalization scheme, which we have chosen such that the anomaly is proportional to $d_{ABC}$. Choosing a different scheme would change the coefficients in the  GCS term.}. Quantum anomalies can cancel the terms in the variation (\ref{delLambdaSf1}) that are proportional to the symmetric part $C_{AB,C}^{(s)}$, provided we have $d_{ABC}=C_{AB,C}^{(s)}$. This is the Green-Schwarz mechanism. Thus, putting everything together, if we have
\begin{equation}
C_{AB,C}=d_{ABC}+C_{AB,C}^{\rm (CS)}, \label{CdC}
\end{equation}
the variation of the GCS term and the quantum anomaly together cancel the variation (\ref{delLambdaSf1}).

In \cite{Anastasopoulos:2006cz}, it was recently argued that
GCS terms (together with anomalous chiral fermions) are generic features of orientifold 
compactifications with intersecting D-branes \cite{ISB}, and that GCS terms may lead to interesting phenomenological signatures for variants of $Z^{\prime}$-bosons (for more details on these and similar models see, e.g., \cite{Kumar:2007zza}). Other possible higher-dimensional origins of GCS terms are described in \cite{Andrianopoli:2004sv,Gunaydin:2005bf}. They also play an important r\^{o}le in the manifestly symplectic formulation of gauged supergravities of \cite{deWit:2005ub}.

Given these interesting applications, it is important to explore to what extent the above mechanisms can can be reconciled with supersymmetry. Quite surprisingly, in the context of $\mathcal{N}=1$ supersymmetry, this has been
fully understood only very recently in \cite{DeRydt:2007vg}, which generalizes earlier work \cite{Andrianopoli:2004sv} on GCS terms in \emph{global} $\mathcal{N}=1$ supersymmetry \emph{without} quantum anomalies.  It should be emphasized that the case of $\mathcal{N}=1$ supersymmetry is qualitatively different from the analogous treatments with extended supersymmetry (as e.g. \cite{Sugra} and the general treatments \cite{deWit:1987ph,deWit:2005ub} that mimic the situation of extended supergravity theories), as the latter theories cannot have chiral gauge interactions and hence no quantum anomalies. 

In the next two sections, I will describe the general implementation
of gauged axionic shift symmetries, GCS terms and quantum anomalies in theories with global and local
$\mathcal{N}=1$ supersymmetry, following the more detailed exposition \cite{DeRydt:2007vg}. In section 4, I then clarify the r\^{o}le of eqs. (\ref{NonAbelianCident}) and (\ref{CdC}) as strong constraints on the allowed quantum anomalies for different types of gauge groups.

\section{Gauge invariance of the fermionic terms}
The action (\ref{L1f}) in $\mathcal{N}=1$ supersymmetry can be obtained from a superspace integral
 \begin{equation}
  S_f =\int \rmd^4 x \rmd^2 \theta\, f_{AB}(X) W_\alpha ^AW_\beta ^B \varepsilon
  ^{\alpha \beta } +\ c.c.,
 \label{Sf}
\end{equation}
where $W_{\alpha}^{A}$ denotes the usual super field strength of the vector superfields $V^{A}$ that contain the vector fields $W_{\mu}^{A}$, and $X^{i}$
are the chiral superfields. This superspace integral contains the fermionic term
\begin{equation}
\ft14 \rmi
(\mathcal{D}_{\mu} \Im f_{AB}) {\bar{\lambda}}^{A}\gamma^{5} \gamma^{\mu}
\lambda^{B} \,,\label{Sfkin}
\end{equation}
where 
\begin{equation}
\label{derivkinfunction} {\cal D}_\mu f_{AB} = \partial_\mu f_{AB} - 2
W_\mu^C f_{C(A}{}^D f_{B)D}\,.
\end{equation}
Note that (\ref{derivkinfunction}) is a fully gauge covariant derivative only if
$C_{AB,C}=0$ and $f_{AB}$ transforms as in (\ref{conditionfinvC1}). On the other hand, if $C_{AB,C}\neq 0$ and $f_{AB}$ transforms as in (\ref{conditionfinvCb}),
the derivative 
(\ref{derivkinfunction}) is no longer covariant, and the fermionic term
(\ref{Sfkin}) transforms by a shift proportional to $C_{AB,C}$ under gauge transformations. In order to get rid of this term, one has to covariantize, by hand, the derivative (\ref{derivkinfunction}) also with respect to the axionic  shifts:
\begin{equation}
\mathcal{D}_{\mu} f_{AB} \longrightarrow
\hat{\cal D}_\mu f_{AB} \equiv \partial_{\mu} f_{AB} -W_{\mu}^{C}\delta_{C} f_{AB}
 = {\cal D}_\mu f_{AB} -  \rmi W_{\mu}^{C} C_{AB,C}\, ,
 \label{fullcovderf}
\end{equation}
which is equivalent to replacing (\ref{Sf}) by
\begin{equation}
  \hat{S}_f= S_f+S_{\rm{extra}}\,, \qquad S_{\rm extra}=\int \rmd^4 x\left(-\ft14 \rmi W_\mu^C
C_{AB,C} \bar\lambda^A \gamma_5\gamma^\mu \lambda^B\right)\,.
 \label{hatSf}
\end{equation}
It should be noted that a superfield formulation that implements this covariantization
%
%
is only known in the case $C_{AB,C}^{(s)}=0$, i.e., when there are no quantum anomalies \cite{Andrianopoli:2004sv}.
In that case, $S_{\rm extra}$ can be combined with $S_{CS}$ to form a superspace integral that is valid in the Wess-Zumino gauge. 

Putting now everything together,   we  have
\begin{equation}
\delta(\Lambda) (\hat{S}_{f}+  S_{CS}) + \int d^4 x \Lambda^{A}\mathcal{A}_{A}=0,
\end{equation}
i.e., a gauge invariant theory, provided that $C_{AB,C}=d_{ABC}+C_{AB,C}^{\rm (CS)}$.

\section{Supersymmetry}
Thus far, we have shown how gauge invariance can be restored in the presence gauged
axionic shift symmetries in general $\mathcal{N}=1$ supersymmetric gauge theories. What we have not yet checked is whether the new action
$\hat{S}_{f}+ S_{CS}$ is also invariant under supersymmetry.
A careful calculation reveals \cite{DeRydt:2007vg}
\begin{eqnarray}
\delta(\epsilon) \left( \hat S_f+S_{\rm CS}\right)=
\int \rmd^4x\, \Re \Big{[} -\ft32 \rmi C_{AB,C}^{\rm (s)}
\bar\epsilon_R \lambda_R^C \bar \lambda^A_L \lambda_L^B\nonumber\\
 - \rmi
C_{AB,C}^{\rm (s)}W_\nu^C
\tilde F^{\mu\nu A} \bar\epsilon_L \gamma_\mu \lambda^B_R 
-\ft{3}{8} C_{AB,C}^{\rm (s)}f_{DE}{}^A
\varepsilon^{\mu\nu\rho\sigma} W_\mu^D W_\nu^E W_\sigma^C \bar\epsilon_L
\gamma_\rho\lambda^B_R\Big{]}\,. \label{totalgaugevar}
\end{eqnarray}
This is not zero, and in fact, it should not be zero. The reason is that the above \emph{classical} action is not \emph{gauge} invariant either. As we are working in the Wess-Zumino gauge, this will also imply a non-invariance under supersymmetry
(for more details, see \cite{DeRydt:2007vg}).
Thus the classical gauge non-invariance triggers a classical supersymmetry non-invariance. However, this is also true for the \emph{quantum} gauge anomaly; it also triggers a supersymmetry anomaly of the quantum effective action, $
\delta(\epsilon)\Gamma[W_{\mu}^{A}]=\int d^4x \bar{\epsilon}\mathcal{A}_{\epsilon}.$
The supersymmetry anomaly has been calculated by Brandt \cite{Brandt:1993vd}, and it is precisely the negative of equation (\ref{totalgaugevar}),
\begin{equation}
\delta(\epsilon) \left( \hat S_f+S_{\rm CS}\right) + \int d^4 x \bar{\epsilon} \mathcal{A}_{\epsilon}=0 \, .  
\end{equation}
Thus,  the entire classical plus quantum theory is indeed supersymmetric. 

As can be verified using the superconformal calculus, all the above also goes through for the supergravity version of this theory \cite{DeRydt:2007vg}.

\section{Gauge group constraints}
In the previous sections, we described how an interplay of Peccei-Quinn terms, generalized Chern-Simons terms and 
quantum anomalies can yield a gauge invariant and supersymmetric theory even though each of these three effects
individually violates gauge and supersymmetry invariance. In this final section\footnote{Part of this section grew out of a discussion with J.~De Rydt and T.~Schmidt.}, we would like to point out that
the underlying cancellation of non-invariances is possible only for certain types of gauge groups. In fact, it is easy to verify that semisimple gauge groups  do \emph{not} fall into this category. One way to see this, is to first note that GCS-terms for a purely semisimple gauge group can always be absorbed into the kinetic term (\ref{L1f}) via a redefinition \cite{deWit:1987ph}
\begin{equation}
f_{AB}^{\prime}=f_{AB}+iZ_{AB}
\end{equation}
with a constant real tensor $Z_{AB}$ satisfying
\begin{equation}
 C_{AB,C}^{\rm (CS)}=2f_{C(A}{}^{D}Z_{B)D}.
\end{equation}
Without loss of generality, we can thus, for semisimple gauge groups, think of $S_{CS}$ as being absorbed into the kinetic term and use $C_{AB,C}^{(m)}=C_{AB,C}^{\rm CS}=0$ in the resulting theory without Chern-Simons terms. Using $C_{AB,C}=C_{AB,C}^{(s)}=d_{ABC}$, the constraint (\ref{NonAbelianCident}) then implies
\begin{equation}
 d_{ABE} f_{CD}{}^{E}=0 . \label{df}
\end{equation}
As a semisimple group has no Abelian ideals, (\ref{df}) implies $d_{ABC}=0$, i.e., a semisimple group must be free
of cubic anomalies, in agreement with the usual assumption.

If the gauge group is of the form Abelian $\times$ semisimple, only the purely semisimple GSC-terms can be absorbed into the kinetic term. In case the other GCS-terms also vanish, we again have equation (\ref{df}), which now implies that
all components of $d_{ABC}$ with at least one semisimple index vanish. In other words,  we can have at most purely Abelian 
cubic anomalies if there are no GCS-terms. If we do allow for GCS-terms, on the other hand, the constraint (\ref{df}) becomes relaxed, and mixed anomalies of the type $\textrm{Abelian } \times \textrm{ semisimple}^2 $ are also allowed, as is consistent with the more explicit examples in \cite{Anastasopoulos:2006cz,DeRydt:2007vg,SUGRAanomalies}. In the standard treatments 
of the Green-Schwarz mechanism, the GCS-terms in this case are  often referred to as counterterms that convert the covariant anomaly to the consistent anomaly (cf. eq. (\ref{delLambdaSf1}) vs. (\ref{gaugeanom})).

\section{Conclusions}
In this talk I have reviewed the general form of theories with Stueckelberg-type shift symmetries, generalized Chern-Simons terms and 
quantum anomalies in the context of $\mathcal{N}=1$ global and local supersymmetry \cite{DeRydt:2007vg}. The simple consistency conditions (\ref{NonAbelianCident}) and (\ref{CdC}) encode all the more specialized consistency conditions and assumptions that are frequently found in the literature and put them on an equal footing.
It would be interesting to further study the consequences of this work both with regard to its possible phenomenological implications as well as 
its relevance for string compactifications with background fluxes and intersecting branes.

%


\section*{Acknowledgments}
It is a pleasure to thank J.~De Rydt, J.~Rosseel, T.~T.~Schmidt and A.~Van Proeyen for the collaboration on \cite{DeRydt:2007vg} 
and the K.U. Leuven for 
hospitality during parts of this work. The work of M.Z. is supported by the German Research Foundation (DFG) within the Emmy-Noether Program (Grant number ZA 279/1-2).



\begin{thebibliography}{10}




\bibitem{DeRydt:2007vg}
  J.~De Rydt, J.~Rosseel, T.~T.~Schmidt, A.~Van Proeyen and M.~Zagermann,
  {\it Class.\ Quant.\ Grav.\ }  {\bf 24}, 5201 (2007)
  [arXiv:0705.4216 [hep-th]].

\bibitem{Sugra}
  B.~de Wit, P.~G.~Lauwers and A.~Van Proeyen,
  {\it Nucl.\ Phys.\  B} {\bf 255} 569 (1985).


\bibitem{deWit:1987ph}
  B.~de Wit, C.~M.~Hull and M.~Rocek,
  {\it Phys.\ Lett.\  B} {\bf 184} 233 (1987).



\bibitem{Andrianopoli:2004sv}
  L.~Andrianopoli, S.~Ferrara and M.~A.~Lledo,
  {\it JHEP} {\bf 0404}, 005 (2004)
  [arXiv:hep-th/0402142].

\bibitem{Anastasopoulos:2006cz}
  P.~Anastasopoulos, M.~Bianchi, E.~Dudas and E.~Kiritsis,
  {\it JHEP} {\bf 0611}, 057 (2006)
  [arXiv:hep-th/0605225];
  P.~Anastasopoulos,
  Fortsch.\ Phys.\  {\bf 55}, 633 (2007)
  [arXiv:hep-th/0701114].
  
\bibitem{ISB}
\emph{E.g.,}  
  C.~Angelantonj and A.~Sagnotti,
  Phys.\ Rept.\  {\bf 371} (2002) 1
  [Erratum-ibid.\  {\bf 376} (2003) 339]
  [arXiv:hep-th/0204089];
A.~M.~Uranga,
  Class.\ Quant.\ Grav.\  {\bf 20}, S373 (2003)
  [arXiv:hep-th/0301032];
  F.~G.~Marchesano Buznego,
  arXiv:hep-th/0307252;
  D.~L\"{u}st,
  Class.\ Quant.\ Grav.\  {\bf 21}, S1399 (2004)
  [arXiv:hep-th/0401156];
  R.~Blumenhagen, M.~Cvetic, P.~Langacker and G.~Shiu,
  Ann.\ Rev.\ Nucl.\ Part.\ Sci.\  {\bf 55}, 71 (2005)
  [arXiv:hep-th/0502005];
  R.~Blumenhagen, B.~K\"{o}rs, D.~L\"{u}st and S.~Stieberger,
  Phys.\ Rept.\  {\bf 445}, 1 (2007)
  [arXiv:hep-th/0610327].

  
\bibitem{Kumar:2007zza}
  D.~M.~Ghilencea, L.~E.~Ibanez, N.~Irges and F.~Quevedo,
  JHEP {\bf 0208}, 016 (2002)
  [arXiv:hep-ph/0205083];
  B.~K\"{o}rs and P.~Nath,
  Phys.\ Lett.\  B {\bf 586}, 366 (2004)
  [arXiv:hep-ph/0402047];
  B.~K\"{o}rs and P.~Nath,
  JHEP {\bf 0412}, 005 (2004)
  [arXiv:hep-ph/0406167];
  B.~K\"{o}rs and P.~Nath,
  JHEP {\bf 0507}, 069 (2005)
  [arXiv:hep-ph/0503208];
  C.~Coriano', N.~Irges and E.~Kiritsis,
  Nucl.\ Phys.\  B {\bf 746} (2006) 77
  [arXiv:hep-ph/0510332];
  D.~Feldman, Z.~Liu and P.~Nath,
  Phys.\ Rev.\ Lett.\  {\bf 97}, 021801 (2006)
  [arXiv:hep-ph/0603039];
  D.~Feldman, B.~K\"{o}rs and P.~Nath,
  Phys.\ Rev.\  D {\bf 75}, 023503 (2007)
  [arXiv:hep-ph/0610133];
  C.~Coriano, N.~Irges and S.~Morelli,
  JHEP {\bf 0707} (2007) 008
  [arXiv:hep-ph/0701010];
  D.~Feldman, Z.~Liu and P.~Nath,
   ``The Stueckelberg Z' extension with kinetic mixing and milli-charged dark
  Phys.\ Rev.\  D {\bf 75}, 115001 (2007)
  [arXiv:hep-ph/0702123];
  C.~Coriano, N.~Irges and S.~Morelli,
  Nucl.\ Phys.\  B {\bf 789} (2008) 133
  [arXiv:hep-ph/0703127];
  J.~Kumar, A.~Rajaraman and J.~D.~Wells,
  arXiv:0707.3488 [hep-ph];
  I.~Antoniadis, A.~Boyarsky and O.~Ruchayskiy,
  arXiv:0708.3001 [hep-ph];
R.~Armillis, C.~Coriano and M.~Guzzi,
  arXiv:0709.2111 [hep-ph]; 
  K.~m.~Cheung and T.~C.~Yuan,
  arXiv:0710.2005 [hep-ph];
  R.~Armillis, C.~Coriano and M.~Guzzi,
  JHEP {\bf 0805} (2008) 015
  [arXiv:0711.3424 [hep-ph]];
  J.~A.~Harvey, C.~T.~Hill and R.~J.~Hill,
  arXiv:0712.1230 [hep-th];
  P.~Langacker,
  arXiv:0801.1345 [hep-ph].
  
  
\bibitem{Gunaydin:2005bf}
  M.~G\"{u}naydin, S.~McReynolds and M.~Zagermann,
  {\it JHEP} {\bf 0601}, 168 (2006)
  [arXiv:hep-th/0511025].

\bibitem{deWit:2005ub}
  B.~de Wit, H.~Samtleben and M.~Trigiante,
 {\it JHEP} {\bf 0509}, 016 (2005)
  [arXiv:hep-th/0507289].


\bibitem{Brandt:1993vd}
  F.~Brandt,
 {\it Class.\ Quant.\ Grav.\ }  {\bf 11}, 849 (1994) [arXiv:hep-th/9306054]; 
 {\it Annals Phys.\ }  {\bf 259}, 253 (1997)
  [arXiv:hep-th/9609192].
  
\bibitem{SUGRAanomalies}
  D.~Z.~Freedman and B.~K\"{o}rs,
  JHEP {\bf 0611}, 067 (2006)
  [arXiv:hep-th/0509217];
  H.~Elvang, D.~Z.~Freedman and B.~K\"{o}rs,
  JHEP {\bf 0611}, 068 (2006)
  [arXiv:hep-th/0606012].

  

  

  
  
  
  
  
  
  
  
  
  
%
%
%

\end{thebibliography}
\end{document}